\documentclass{Interspeech2024}
\newif\ifinterspeechfinal
\interspeechfinaltrue
\usepackage{algorithm,color,cite}
\usepackage{algorithmic}
\usepackage{multirow}
\usepackage{multicol}
\usepackage{url}
\usepackage{hyperref}
\title{LDM-SVC: Latent Diffusion Model Based Zero-Shot Any-to-Any Singing Voice Conversion with Singer Guidance}

\name[affiliation={1,2}]{Shihao}{Chen}
\name[affiliation={2}]{Yu}{Gu}
\name[affiliation={1}]{Jie}{Zhang}
\name[affiliation={2}]{Na}{Li}
\name[affiliation={2}]{Rilin}{Chen}
\name[affiliation={1}]{Liping}{Chen}
\name[affiliation={1}]{Lirong}{Dai}

\address{
  $^{1}$NERC-SLIP, University of Science and Technology of China (USTC), China \\
  $^{2}$ Tencent AI Lab
  \thanks{
  This work was done at Tencent AI Lab as an internship by Shihao Chen.
  This work was supported by the National Natural Science Foundation of China (62101523), Hefei Municipal Natural Science Foundation (2022012) and USTC Research Funds of the Double First-Class Initiative (YD2100002008). (Correspondence: jzhang6@ustc.edu.cn)
}
}
\email{shchen16@mail.ustc.edu.cn, colinygu@tencent.com, jzhang6@ustc.edu.cn}

\keywords{Singing voice conversion, latent diffusion model, variational autoencoder, classifier-free guidance}

\begin{document}

\maketitle

\begin{abstract} 
    Any-to-any singing voice conversion (SVC) is an interesting audio editing technique, aiming to convert the singing voice of one singer into that of another, given only a few seconds of singing data. However, during the conversion process, the issue of timbre leakage is inevitable: the converted singing voice still sounds like the original singer's voice. To tackle this, we propose a latent diffusion model for SVC (LDM-SVC)  in this work, which attempts to perform SVC in the latent space using an LDM. We pretrain a variational autoencoder structure using the noted open-source So-VITS-SVC project based on the VITS framework, which is then used for the LDM training. Besides, we propose a singer guidance training method based on  classifier-free guidance to further suppress the timbre of the original singer. Experimental results show the superiority of  the proposed method over previous works in both subjective and objective evaluations of timbre similarity. 
\end{abstract}

\section{Introduction}

Singing Voice Conversion (SVC) is a popular audio editing technique that aims to change the singing voice of one singer to mimic another.
Different from singing voice synthesis which requires well-designed musical note inputs \cite{bytesing,cui2024sifisinger}, 
this technique allows users to customize their favorite singers performing any songs just given  corresponding recorded songs sung by other singers.
 Unlike the many-to-many or many-to-one scenarios,
 the any-to-any SVC is much more challenging, which demands the model to perform conversion for any target singer who were not included in the training set by solely a short snippet of reference singing voice that even lasts for few seconds.

The main challenge of SVC is to separate and reassemble the singer's unique vocal timbre from the content and melody of songs. 
Similarly to voice conversion,  mainstreaming SVC systems also follows a recognition-synthesis scheme as a typical two-stage process. In the first stage, singer-independent features such as phonetic posteriorgrams (PPG) \cite{sun2016phonetic, polyak2020unsupervised,liu2021fastsvc,liu2021diffsvc,li2021ppg} from an ASR model and self-supervised learning (SSL) representations \cite{jayashankar2023self,zhou2023vits} trained on large amounts of unlabeled speech data are used to encode audio. 
These representations serve as intermediary for SVC, which can effectively extract content and  semantic information from waveforms.
In the second stage, acoustic models are involved to generate the target audio or acoustic features 
from these immediate representations.
 Various generative models have been employed for SVC decoding, including autoregressive models \cite{nachmani2019unsupervised,deng2020pitchnet,zhang2020durian,takahashi2021hierarchical}, generative adversarial networks (GANs) \cite{polyak2020unsupervised,liu2021fastsvc,zhou2022hifi}, variational autoencoder (VAE) \cite{luo2020singing} and diffusion models \cite{liu2021diffsvc}.
Despite of naturalness, sound quality and intonation accuracy of converted singing voice have largely improved by above different SVC models, the timbre leakage problems remain serious, especially for the challenges of SVC cross different genders. This is primarily due to PPG and SSL features containing not only content information but also some timbre information of the original singer.To alleviate the timbre leakage,  many works such as So-VITS-SVC\footnote{\url{https://github.com/PlayVoice/so-vits-svc-5.0/tree/bigvgan-mix-v2}} involved information perturbation by directly adding white noises on 
 hidden features or acoustic features. 
 However such white noises were totally independent with singer information and the information perturbation modules were not trainable and optimized in the network training stage and directly adding noise on acoustic features may also lead in pronunciation and quality distortion.

 \begin{figure*}[t]
  \centering
  \includegraphics[width=\textwidth]{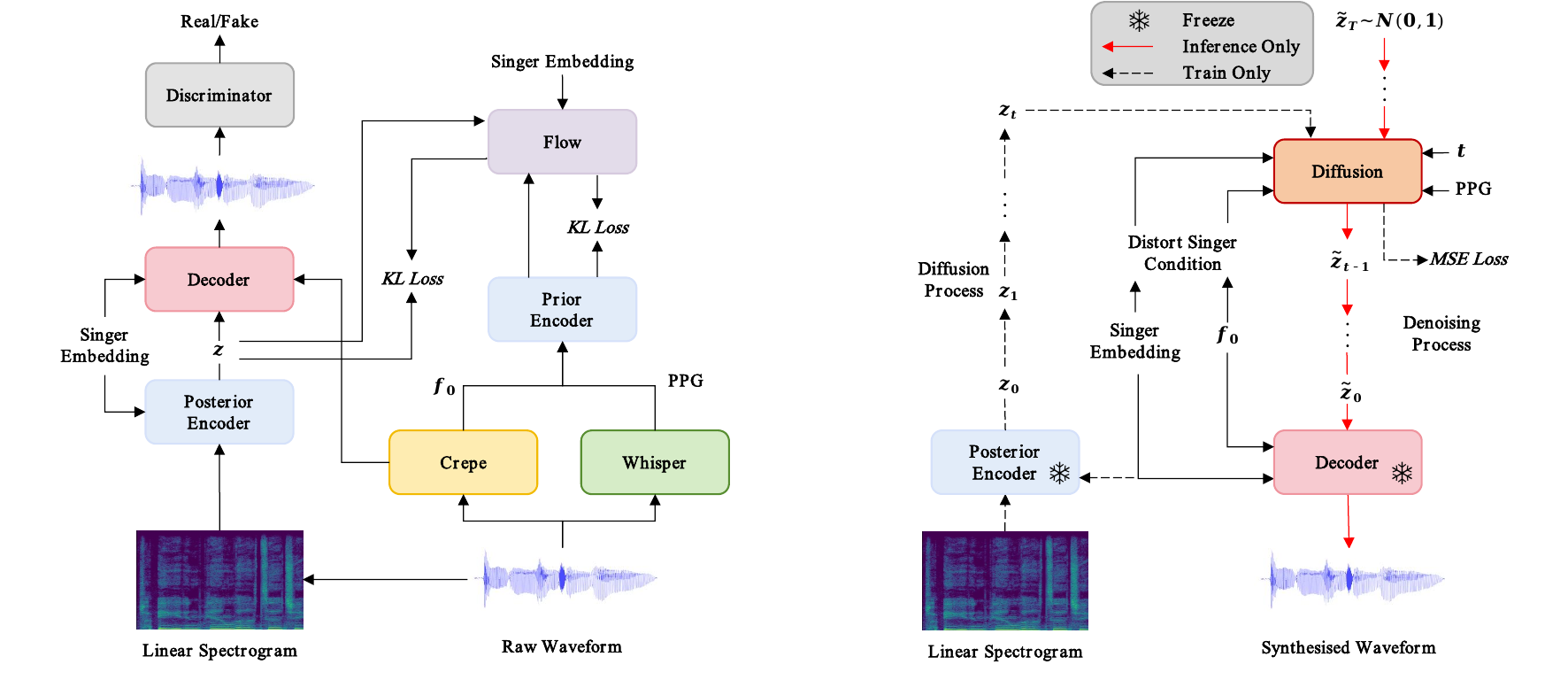}
  \caption{Left: Pre-training procedure of So-VITS-SVC; Right: Training procedure of LDM-SVC.}
  \label{ldm-svc}
\end{figure*}

Recently, Latent Diffusion Model (LDM) based systems have shown a great success in image generation from text such as Stable Diffusion \cite{rombach2022high} and high-quality sound generation from text such as Tango~\cite{ghosal2023text} and  AudioLDM~\cite{liu2023audioldm, audioldm2}, which performed forward and denoising diffusion processes on the   hidden spaces rather 
than acoustic features like other models \cite{liu2022diffsinger, popov2021diffusion}. Motivated by these models,  we present LDM-SVC, a novel any-to-any SVC method which reconstructs the waveform directly from the latent representation in an end-to-end latent diffusion manner. Unlike DiffSVC \cite{liu2021diffsvc} employed on mel-spectrograms and requiring an additional vocoder, we conduct the diffusion model on the hidden space from a pre-trained VAE model using  So-VITS-SVC and directly
 utilize the predicted latent representation to generate the waveforms by the VAE decoder.
To address the timbre leakage issue, we regard the LDM forward process as a information perturbation process in those both progressively adding noise to decouple the singer timbre from 
content and melody. Different from the aforementioned methods which simply add white noise on waverforms, such information perturbation module is trainable and conditioned on singer information.
To better decouple singer information, a singer guidance training mechanism is explored, which is inspired by the  classifier-free method \cite{ho2022classifier} in image generation when training the conditional and unconditional diffusion model at the same time. Comparing with
many state-of-the-art SVC models, both subjective and objective experimental results indicate that our proposed method can achieve greater timbre similarity in any-to-any SVC tasks for both seen singer conversion and unseen singer conversion scenarios and better singing naturalness.

The rest of this paper is organized as follows. Section 2 outlines the proposed LDM-SVC method. Experiments are presented in Section 3. Finally, Section 4 concludes this work.

\section{Proposed Method}
\subsection{VAE Pretraining}
\label{vae pretrain}
We pretrain the VAE model using So-VITS-SVC, which is a VAE based any-to-any SVC model following the VITS framework \cite{kim2021conditional} and consists of three key components: posterior encoder, prior encoder and decoder as depicted in Figure 1. 
The posterior encoder $\mathcal{E}(\cdot)$ composed of non-causal WaveNet \cite{oord2016wavenet} residual blocks models the posterior distribution $p(z|y,e)$ of the hidden representation $z=\mathcal{E}(y,e)$ from the linear spectrograms  $y$ generated from the original singing waveforms  where singer embedding $e$ is extracted by an additional speaker verification model.
The prior encoder is implemented using a multi-layer Transformer \cite{vaswani2017attention}. Given the PPG and fundamental frequency (F0) denoted as $x$ and $f_0$ respectively, the prior encoder estimates the prior distribution $p(z|x,f_0,e)$ with the target singer's timbre and the flow. 
To bridge the distribution between the prior encoder and posterior encoder, normalizing flow with speaker-normalized affine couplin (SNAC) \cite{choi2022snac} layers is exploited to perform an invertible transformation of a simple distribution into a more complex one. 
The BigVGAN-based decoder \cite{lee2022bigvgan} $\mathcal{D}(\cdot)$ generates the singing waveform from the latent representation $z$ using a neural source filter (NSF) scheme \cite{wang2019neural} with F0 to enhance voice reconstruction quality. 

After training the So-VITS-SVC system, we retain only the posterior encoder and decoder. The posterior encoder compresses the linear spectrogram to generate the latent representation, used as the prediction target for the diffusion model in LDM-SVC training. During inference, the latent representation is predicted from Gaussian White Noise by the denoising process, and the waveform is generated via the decoder.
 
\subsection{Latent Diffusion}
LDM is  adopted as the probabilistic models that fit the hidden distribution by denoising on data latent space from pretrained VAE model. We use the Denoising Diffusion Probabilistic Models (DDPM) method \cite{ho2020denoising} to train the diffusion model, which consists of forward and denoising processes. Initially, we pretrain an SVC model using the So-VITS-SVC framework and combine its posterior encoder and decoder to form a VAE (see the right part of Figure 1). During the LDM training, the singer's timbre $e$ and the linear spectrogram of the singing voice $y$ are used as inputs to the posterior encoder $\mathcal{E}(\cdot)$, yielding the latent variable $z_0=z=\mathcal{E}(y,e)$. 
In the forward process, the original data distribution is transformed into a standard Gaussian distribution by gradually adding noise according to a fixed schedule $\beta_1,\dots,\beta_T$. Here, $T$ represents the total timesteps. The transition from $z_{0}$ to $z_t$ follows a Markov chain, where the conditional distribution $q(z_t|z_{t-1})$ is defined as a Gaussian distribution: $q(z_t|z_{t-1}) = \mathcal{N}(z_t;\sqrt{1-\beta_t}z_{t-1}, {\beta_t}\mathbf{I})$.

The denoising process, parameterized by $\theta$, acts as a denoising function, eliminating the added noise and restoring the original data structure. This denoising distribution, $p_{\theta}(z_{t-1}|z_t)$, is modeled as a conditional Gaussian distribution. By using this parameterized denoising process, we iteratively sample the target data $z_0$ from a Gaussian noise for $t = T, T-1, \ldots, 1$. In each iteration, $z_{t-1}$ is sampled according to $p_{\theta}(z_{t-1}|z_t)$. The LDM model receives $z_t$ and $t$ as inputs, complemented by conditional inputs such as the singer's timbre $e$, fundamental frequency $f_0$, and PPG $x$. To put it simply, in the denoising procedure the calculation of $z_{t-1}$ is given by
\begin{equation}z_{t-1} = \frac{1}{\sqrt{\alpha_t}} \left( z_t - \frac{1-\alpha_t}{\sqrt{1-\bar{\alpha}_t}} \epsilon_{\theta}(z_t, t, x, f_0, e) \right) + \sigma_t \epsilon,
\end{equation}
where $\epsilon \sim \mathcal{N}(0, I)$ represents Gaussian white noise, $\alpha_t = 1 - \beta_t$, and $\bar{\alpha}_t = \prod_{s=1}^t \alpha_s$. For a more detailed derivation of (1), please refer to \cite{ho2020denoising}.
The configuration of the diffusion model we use aligns with that in DiffSVC.  
The training loss of the latent diffusion model $\epsilon_\theta$ is defined as the
mean squared error (MSE) in the noise space:
\begin{equation}
    \mathcal{L}_{LDM}=||\epsilon-\epsilon_\theta(z_t, t, x, f_0, e)||_2^2.
\end{equation}

During inference, the singer's timbre and F0 are replaced with the target singer's attributes, defined as $e_{tar}$ and $f_0$, while the source singer's PPG $x_{src}$ is used as a condition. We sample a random Gaussian White Noise in the denoising process to obtain latent variable $z_0$. Consequently, $z_0$, $e_{tar}$, and $f_0$ are fed into the pretrained decoder $\mathcal{D}(\cdot)$ to generate the audio waveform. The use of LDM can alleviate the inconsistency caused by training-testing mismatches in So-VITS-SVC.
\begin{algorithm}[t]
\caption{Training procedure of LDM-SVC.}
\begin{algorithmic}[1]
\label{alg:training}
\REQUIRE Conversion model $\epsilon_\theta(\cdot)$; training set $D_{train} = \{(x,{f_0},e,y)\}_{m=1}^M$; pretrained So-VITS-SVC posterior encoder $\mathcal{E(\cdot)}$; distortion probability $p_{uncond}$; $N_{iter}$ iterations.\FOR{$i=1,2,...,N_{iter}$}
\STATE Sample $(x, f_0, e, y)$ from $D_{train}$;
\STATE $z_0=\mathcal{E}(y,e)$;
\STATE $e,f_0 \leftarrow \varnothing$ with probability $p_{uncond}$;
\STATE $\epsilon\sim\mathcal{N}(0,I)$;
\STATE Sample $t\sim$ Uniform$(\{1, \cdots, T\})$;
\STATE Take gradient descent step on \\ \quad$\nabla_\theta||\epsilon-\epsilon_\theta(\sqrt{\bar\alpha_t}z_0+\sqrt{1-\bar\alpha_t}\epsilon, t, x, f_0, e)||_2^2$;
\ENDFOR
\RETURN $\epsilon_\theta(\cdot)$;
\end{algorithmic}
\end{algorithm}

\begin{algorithm}[t]
\caption{Inference procedure of LDM-SVC.}
\begin{algorithmic}[1]
\label{alg:infer}
\REQUIRE Trained conversion model $\epsilon_\theta(\cdot)$; source singer 
PPG $x_{src}$; target singer embedding $e_{tar}$; modified $f_0$; pretrained So-VITS-SVC decoder $\mathcal{D}(\cdot)$; guidance weight $w$.\STATE Sample $z_T\sim\mathcal{N}(0,I)$;
\FOR{$t=T,T-1,...,1$}
\STATE $\epsilon_t=(1+w)\epsilon_\theta(z_t, t, x, f_0, e_{tar})-w\epsilon_\theta(z_t, t, x, \varnothing, \varnothing)$;\STATE $\epsilon\sim\mathcal{N}(0,I)$ if $t>1$ else $\epsilon=0$;
\STATE $z_{t-1} = \frac{1}{\sqrt{\alpha_t}}(z_t - \frac{1-\alpha_t}{\sqrt{1-\bar{\alpha}_t}}\epsilon_t)+\sigma_t \epsilon$;
\ENDFOR
\RETURN $\mathcal{D}(z_0,f_0,e_{tar})$;
\end{algorithmic}
\end{algorithm}
\subsection{Singer Guidance}
 As in Figure 2, we employ the speaker condition layer normalization (SCLN) \cite{wu2021cross} to normalize the PPG feature. Additionally, we use classifier-free guidance \cite{ho2022classifier} to train the model. Specifically, we expect the model to predict $p(z|x)$ via the score estimator $\epsilon_\theta(z_t, t, x)$ obtained from unconditional diffusion, while simultaneously predicting $p(z|x, e, f_0)$ through the score estimator $\epsilon_\theta(z_t, t, x, e, f_0)$ obtained from the conditional diffusion, as summarized in Algorithm \ref{alg:training}. During model inference, we perform two inferences. In the first inference, we input all conditions normally. In the second, we set $e_{tar}$ and $f_0$ to an empty set.  Finally, we perform sampling using the following linear combination of the conditional and unconditional score estimates with a guidance weight $w$ as shown in Algorithm \ref{alg:infer}:
\begin{equation}
    \epsilon_t=(1+w)\epsilon_\theta(z_t, t, x, f_0, e_{tar})-w\epsilon_\theta(z_t, t, x, \varnothing, \varnothing).
\end{equation}
Based on the linear combination of predictions from both the conditional and unconditional models, we can thus more effectively reduce the timbre information of the source singer.
\begin{figure}[]
  \centering
  \includegraphics[width=\textwidth/2]{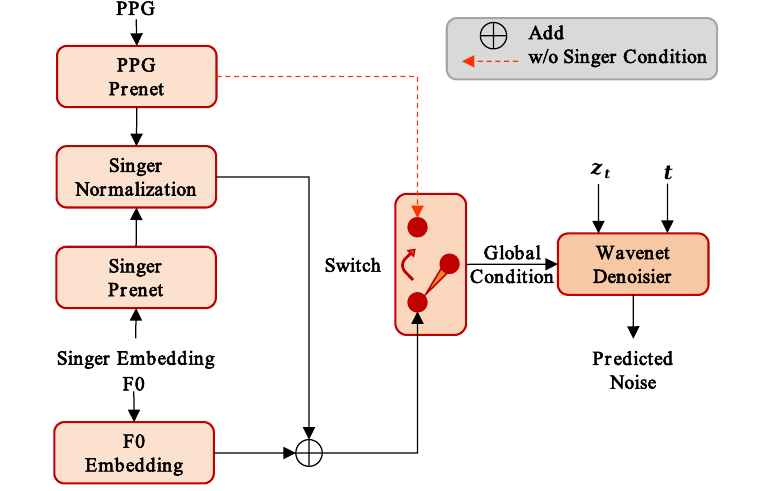}
  \caption{Singer guidance using a latent diffusion model.}
  \label{guidance}
\end{figure}

\section{Experiment}
\begin{table*}[ht]
\caption{Subjective indicators (SMOS, NMOS) and objective indicators (SSIM, FPC) under various method.}
\renewcommand{\arraystretch}{1}
\setlength\tabcolsep{2pt} 
\centering
\resizebox{\textwidth}{!}{%
\begin{tabular}{@{}l|cccccccc@{}}
\toprule
\multirow{2.5}{*}{Method} & \multicolumn{4}{c|}{Seen Singer}                          & \multicolumn{4}{c}{Unseen Singer}      \\ \cmidrule(l){2-9} 
 &
  \multicolumn{1}{c}{NMOS} &
  \multicolumn{1}{c}{SMOS} &
  \multicolumn{1}{c}{SSIM} &
  \multicolumn{1}{c|}{FPC} &
  \multicolumn{1}{c}{NMOS} &
  \multicolumn{1}{c}{SMOS} &
  \multicolumn{1}{c}{SSIM} &
  \multicolumn{1}{c}{FPC} \\ \midrule
FastSVC      &  $2.195\pm0.053$  & $2.041\pm0.029$ & $0.255$ & \multicolumn{1}{c|}{$0.734$}   &$2.245\pm0.073$ & $1.788\pm0.066$& $0.287$ & $0.696$ \\
DiffSVC      & $3.950\pm0.087$  & $3.681\pm0.115$& $0.593$ & \multicolumn{1}{c|}{$0.944$}  &$3.720\pm0.108$ & $3.444\pm0.127$ & $0.586$ & $0.942$ \\
So-VITS-SVC   & $3.985\pm0.082$  & $3.578\pm0.112$& $0.639$ & \multicolumn{1}{c|}{$\bf{0.946}$} &$3.906\pm0.095$ & $3.416\pm0.124$ & $0.600$ & $0.942$ \\
 LDM-SVC-w/o-SD &$4.067\pm0.075$  & 
$\bf{3.950\pm0.116}$ & $0.680$ & \multicolumn{1}{c|}{$0.945$}  &$3.937\pm0.093$ & $3.650\pm0.139$ & $0.640$ & $0.940$\\
 LDM-SVC   &  $\bf{4.125\pm0.080}$ & $3.947\pm0.114$  & $\bf{0.702}$ &  \multicolumn{1}{c|}{$0.945$}& $\bf{4.025\pm0.100}$ & $\bf{3.841\pm0.127}$ & $\bf{0.663}$& $\bf{0.942}$ 
\\
 \bottomrule
\end{tabular}
}
\label{result_eer}
\end{table*}

\begin{figure*}[ht]
  \centering
  \includegraphics[width=\textwidth]{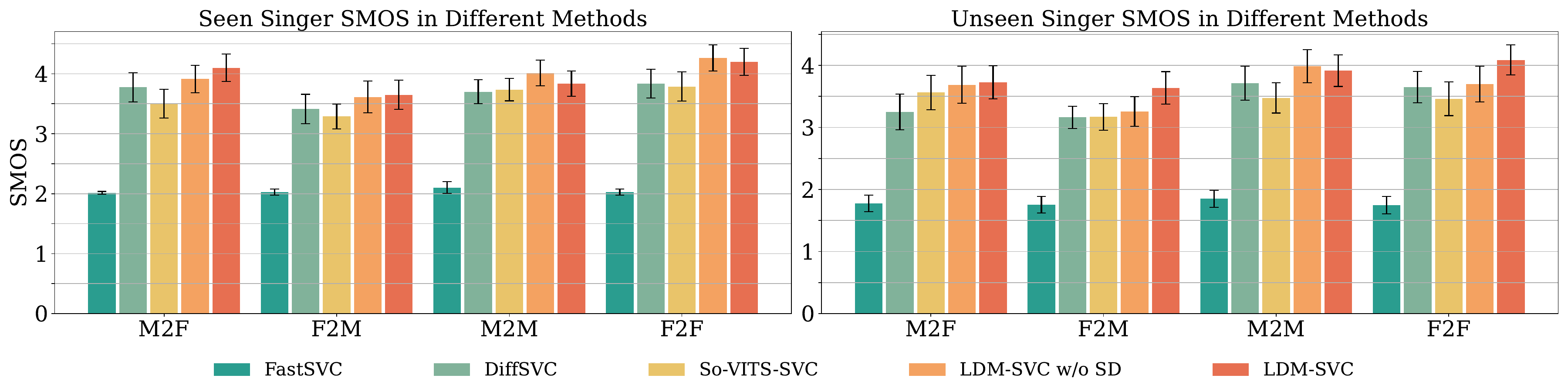}
  \caption{Detailed SMOS for seen and unseen scenarios, including M2M, M2F, F2M and F2F.}
  \label{gender}
\end{figure*}

\subsection{Experiment Setup}
To better accomplish the any-to-any SVC task, we choose the OpenSinger \cite{huang2021multi} dataset, which contains a large number of Chinese singers. The dataset comprises 74 singers, consisting of 27 males and 47 females, and amounts to a total of 52 hours of recordings. 
To evaluate the conversion efficacy of the proposed system, we test two cases separately on whether  the target singers are included in the training dataset (seen singers) or not (unseen singers).
In the test set, 4 male and 4 female singers are randomly selected as  the unseen singers. For the case of seen singers, 8 sentences from each remaining 66 singers are chosen 
for testing and the rest of songs are used for training.

We train FastSVC, DiffSVC and So-VITS-SVC models for comparison, where the So-VITS-SVC model is also the VAE used in our LDM method. To ensure a fair comparison, the PPG for each model is extracted using Whisper \cite{radford2023robust}, while F0 is extracted using CREPE  \cite{kim2018crepe}. The singer embeddings are obtained from the pre-trained model of the CAM++ \cite{wang2023cam++},  a speaker verification open-source project on modelscope\footnote{\url{https://modelscope.cn/models/iic/speech_campplus_sv_zh-cn_16k-common/summary}}. In experiments, all audio files are resampled to 32kHz. For F0 embedding in the diffusion condition, we first quantize the Log-F0 features into 256 bins and then go through a melody embedding lookup table. During inference, we re-edit the F0 of the original singer. Specifically, we calculate the mean F0 values of the target singer and source singer in the voice segment, denoted as ${\rm mean}({f_0}^{src}_v)$ and ${\rm mean}({f_0}^{tar}_v)$ respectively. Then, we multiply  ${f_0}^{src}$ with the ratio of these mean values to obtain the modified ${f_0}$ as ${f_0}^{src} \times {{\rm mean}({f_0}^{tar}_v)}/{{\rm mean}({f_0}^{src}_v)}$. For the singer guidance method, we set the distortion probability $p_{uncond}$ during training to 0.1, and the guidance weight $w$ during inference to 0.3 similarly to \cite{ho2022classifier}. The total number of diffusion steps is 100 (i.e., T = 100). The noise schedule, represented as $\beta$, is configured to be linearly distributed, ranging from a very small value of $1 \times 10^{-4}$ up to $0.06$. This setup follows the approach used in the DiffSVC model. In our DiffSVC experiment, to ensure a fair comparison, we use BigVGAN with NSF as the unified vocoder in accordance with the structure adopted in So-VITS-SVC.
Audio samples are available at \url{https://sounddemos.github.io/ldm-svc}.

\subsection{Evaluation Metrics}
  To construct the conversion trials, we separately establish pairs of conversions for both seen and unseen scenarios and a cross-validation strategy  is employed  to conduct audio clips where each singer provides the vocal content information as source and other singers provide timbre information as targets.

  We evaluate these converted audios using F0 Pearson correlation (FPC) and Singer Similarity (SSIM), with SSIM calculated via cosine similarity using speaker embedding from the Automatic Speaker Verification model. For subjective evaluation, we randomly pick 40 audio samples from each model in both seen and unseen senarios. These include  Male-to-Male (M2M), Male-to-Female (M2F), Female-to-Male (F2M) and Female-to-Female (F2F), and each has 10 conversions. 
 We conduct a 5-point Mean Opinion Score (1-bad, 2-poor, 3-fair, 4-good, 5-excellent) test using these randomly selected audio clips. We invite 10 volunteers to participate in the listening test, where both the Naturalness Mean Opinion Score (NMOS) and Similarity Mean Opinion Score (SMOS) are evaluated.
 
\subsection{Experimental Results}
Results for seen and unseen singer cases are shown in Table \ref{result_eer}. Overall, the SMOS, NMOS, SSIM and FPC in the seen scenario are generally better than those in the unseen scenario, which is consistent with common sense. 
In Table 1, the fluency and similarity results of FastSVC are at a relatively low level, indicating a poor applicability to any-to-any conversion, and the generated audio is of poor quality and has no conversion effect.
For the DiffSVC and So-VITS-SVC methods, the scores on NMOS are higher, but the SMOS is relatively low, implying that in case of only providing a few seconds of the target singer's audio, these SVC models may not be able to convert source inputs to the target singer's singing well. Comparing DiffSVC with So-VITS-SVC, it can be observed that the end-to-end generative model So-VITS-SVC produces a better sound quality than the two-stage DiffSVC model. However, the So-VITS-SVC model tends to have more severe timbre leakage.
The results of proposed system LDM-SVC and the ablation system without singer guidance mechanism (LDM-SVC-w/o-SD) are also illustrated in Table 1.
It is clear that both variants have higher SMOS than comparison methods and also have some improvement in NMOS. This discrepancy may be due to the fact that we directly predict $z$ generated by the posterior encoder, whereas during inference with the So-VITS-SVC model, the prior encoder and flow are used to predict $z$ that is approximated by the posterior encoder during training, leading to a mismatch between training and testing.
Comparing the inclusion of singer guidance, the SMOS in the seen and unseen cases are at the same level. However, in the unseen case, the similarity of the model trained with singer guidance is higher than that in the seen case, and the SMOS scores are closer. This reveals that the proposed singer guidance method is effective in the zero-shot scenario. 

The SMOS results for the gender-specific SVC tests are shown in Figure 3. We see that the SMOS scores for all `to male' conversions are significantly lower than those for `to female' cases, which may be due to the imbalance in the dataset between male and female singers. Overall, the issue of timbre leakage in cross-gender conversion is more severe than in same-gender conversions. For example, in the case of M2M the similarity of the model using the singer guidance method is lower than that only using the LDM method, which means that singer guidance in M2M conversions is less effective than others. 
\section{Conclusion}

In this paper, we proposed the LDM-SVC system, which utilizes an LDM for any-to-any singing voice conversion.  We pre-trained a So-VITS-SVC model, whose posterior encoder and decoder were used to construct a VAE to handle the latent representation. It was shown that the proposed LDM-SVC method outperforms existing approaches in terms of timbre similarity in  both seen and unseen singer conversion cases. The designed singer guidance method is beneficial for improving the conversion similarity under zero-shot conditions for unseen singer conversions. In the future, we will consider cross-domain SVC tasks, e.g., converting input speech into singing voice, as a necessity for low-resource scenarios without singing voice data.

\bibliographystyle{IEEEtran}
\bibliography{mybib}

\end{document}